\documentclass[twocolumn,amsmath,amssymb,superscriptaddress]{revtex4}
\usepackage{graphicx}% Include figure files
\usepackage{dcolumn}% Align table columns on decimal point
\usepackage{bm}% bold math

\begin{document}

%\preprint{APS/123-QED}

\title{Radiating black hole solutions in Einstein-Gauss-Bonnet gravity}

\author{Alfredo E. Dominguez}
\email{domingue@fis.uncor.edu} \affiliation{FaMAF, Ciudad
Universitaria, Universidad Nacional de C\'ordoba, (5000),
C\'ordoba, Argentina} \affiliation{Instituto Universitario
Aeron\'autico, Av. Fuerza A\'erea km 6.5. (5000), C\'ordoba,
Argentina}
\author{Emanuel Gallo}
\email{egallo@fis.uncor.edu}
 \affiliation{FaMAF, Ciudad Universitaria, Universidad Nacional de C\'ordoba, (5000), C\'ordoba, Argentina}
\date{\today}
\begin{abstract}

In this paper, we find some new exact solutions to the
Einstein-Gauss-Bonnet equations. First, we prove a theorem which
allows us to find a large family of solutions to the
Einstein-Gauss-Bonnet gravity in $n$-dimensions. This family of
solutions represents dynamic black holes and contains, as particular
cases, not only the recently found Vaidya-Einstein-Gauss-Bonnet
black hole, but also other physical solutions that we think are new,
such as, the Gauss-Bonnet versions of the Bonnor-Vaidya(de
Sitter/anti-de Sitter) solution, a global monopole and the Husain
black holes. We also present a more general version of this theorem
in which less restrictive conditions on the energy-momentum tensor
are imposed. As an application of this theorem, we present the exact
solution describing a black hole radiating a charged null fluid in a
Born-Infeld nonlinear electrodynamics.

\end{abstract}

\pacs{Valid PACS appear here}% PACS, the Physics and Astronomy
%Use showkeys class option if keyword
\maketitle

%\preprint{APS/123-QED}
%\pagenumbering{arabic}

\section{Introduction}

In recent years, there has been a renewed interest in theories of
gravity in higher dimensions. The motivation arises from string
theory. As a possibility, the Einstein-Gauss-Bonnet (EGB) gravity
theory is selected by the low energy limit of the string
theory~\cite{cuerdas1,cuerdas2,cuerdas3}. In this theory appears
corrective terms to Einstein gravity which are quadratic in the
curvature of the space-time. For 4-dimensional gravity, these
terms result in a topological invariant so they will have no
consequences in the field equations of the theory unless a surface
term is involved. However, it was shown in ~\cite{Olea1,Olea2},
that such terms modify the conserved current of the theory in four
dimensions. Moreover, the effect of those Gauss Bonnet terms is
nontrivial for higher dimensions, so the theory of gravity, which
includes Gauss-Bonnet terms, is called Einstein-Gauss-Bonnet (EGB)
gravity. Studies in this area were made in ~\cite{ZW}.

Perhaps black holes are the most striking prediction of any theory
of gravity. This issue was also discussed in EGB gravity. Efforts
were addressed to the understanding of properties of isolated
black holes in equilibrium. In particular, for spherically
symmetric space-time, solutions describing static charged black
holes for both Maxwell and Born-Infeld electrodynamics and other
fields were found in the
literature~\cite{BD,WH1,WH2,WSH1,Giribet}. Thermodynamic
properties of these solutions were also studied
in~\cite{WSH2,Ca1,Ca2}. On physical grounds, one would expect that
an interesting solution should be stable under non-spherically
symmetric perturbations of the state of the black hole represented
by this solution. This subject was investigated by Dotti and
Gleiser in~\cite{DG1,DG2,GD}. Furthermore, the causal structure of
these solutions has been studied in detail by Torii and Maeda in
~\cite{MaedaN,MaedaC}.

However, nature behaves in more complex way: black holes are dynamic
systems, which are seldom in equilibrium. Then, from a physical
point of view, the above cited static solutions should represent the
eventually steady state of dynamic evolution of black holes. This
kind of solutions -- dynamical black holes-- has a twofold value.
First, they allow us to model more realistic physical situations
associated with the black hole dynamic, such as processes of
formation and evaporation of black holes. Second, these solutions
can be used to evaluate the cosmic censorship
hypothesis~\cite{Penrose} which states that naked singularity are
forbidden in physical gravitational collapses.

In the framework of GR, the Vaidya metric~\cite{Vaidya} represents
a radial null fluid which can be used to model dynamic processes
associated with black holes. There exists many possible
generalizations of the Vaidya metric -- see~\cite{Ghosh, Mark} and
references therein--. At this point, it is important to note that
some dynamic solutions describing the collapse of certain kinds of
matter seems to evolve to the formation of naked singularities.
Unfortunately, no much information is available about the
stability of these solutions.

Recently, Dadwood and Ghosh~\cite{Ghosh} have proved a theorem which
is readily seen to generate dynamic solutions of black holes in GR,
by imposing some conditions on energy-momentum tensor. This theorem
is a generalization of a previous one obtained by Salgado in
~\cite{Salgado} and that was generalized for $n$-dimension by Gallo
~\cite{Gallo}.

Unfortunately, in EGB gravity a few solutions representing dynamic
black holes are known. Effectively, a recent Vaidya-type solutions
in the context of EGB gravity were independently obtained by
Kobayashi and Maeda in~\cite{Kobayashi,MaedaN2}. The possibility
that black hole space-times evolve to the formation of naked
singularity were mentioned in ~\cite{MaedaC,MaedaN2}. On the other
hand, Grain, Barrau and Kanti have recently calculated the greybody
factors associated to Einstein-Gauss-Bonnet black holes, which are
needed to describe the evaporation spectrum by Hawking
radiation~\cite{Grain}. Konoplya studied in ~\cite{Kona1} the
quasinormal modes of charged EGB black holes, and in ~\cite{Kona2},
he analyzed the evolution of a scalar field in EGB.

The aim of this work is to prove two similar theorems that were
presented in~\cite{Ghosh} but in the framework of EGB gravity. The
first one is an extension of that theorem for EGB gravity; the
second one is deduced from the first one but relaxing some
conditions on energy-momentum tensor, in order to include more
realistic physical situations. As a consequence of these theorems,
we find the analogous solutions, in EGB gravity, of many well
known solutions for GR (Vaidya, Bonnor-Vaidya, Husain, Global
Monopole). Also, we obtain the Vaidya-type generalization for
static black holes in EGB gravity, such as Born-Infeld solutions.

In section II, we review the EGB gravity and recalled some of the
well known static spherically symmetric black hole solutions. In
section III, we prove the former theorem and we show some
solutions resulting from this theorem. In section IV, we analyzed
the imposition of the energy conditions on these metrics. Finally,
in section V we present a more general version of the latter
theorem and we apply it to obtain the metric of Born-Infeld
dynamic black hole. In the conclusions we discuss possible future
works in this area.

\section{The Einstein-Gauss-Bonnet gravity}

The action which describe Einstein-Gauss-Bonnet gravity coupled
with matter fields reads,
%\begin{widetext}
\begin{equation*}
S = \frac {1}{16\pi } \int d^nx \sqrt {-g}  \left[ R - 2\Lambda +
\alpha (R_{a b c d } R^{a b c d } + \right.
\end{equation*}
\begin{equation*}
\left. + R^2 - 4R_{a b } R^{a b} )\right]+S_{\text{matt}},
\end{equation*}
%\end{widetext}
where $S_{\text{matt}}$ is the action associated to the matter
fields, and $\alpha $ is the Gauss-Bonnet coupling constant
associated in the string models, with the tension of these
strings. This constant introduce a length scale. In fact, the
correction that these theory produce to GR, are noted in short
distance, given by the scale $l=\sqrt{4 \alpha}$.

The equations of motion resulting from $\delta S=0$ are
%\begin{widetext}
\begin{eqnarray*}
8\pi T_{a b } &=& \mathcal{G}_{a b}=G^{(0)}_{a b}+G^{(1)}_{a
b}+G^{(2)}_{a b},
\end{eqnarray*}
where $T_{a b}$ is the energy-momentum tensor, representing the
matter-field distribution resulting from the variation $\delta
S_{\text{matt}}/\delta g^{a b},$ and
\begin{eqnarray*}
G^{(0)}_{a
b}&=&\Lambda g_{a b}  \\
 G^{(1)}_{a
b}&=& R_{a b }-\frac{1}{2}Rg_{a b}\\
G^{(2)}_{a b}&=&-\alpha \left[\frac {1}{2} g_{a b} (R_{ c j e
k}R^{c
 j e k }-4R_{c j }R^{c j }+R^2) \right. -   \\
&-&\left. 2RR_{a b}+4R_{a c}R^{c}_{b}+4R_{c j} R^{c j}_{ \ \ a
b}-2R_{a c j e}R_{b}^{ \ c j e} \right ].
\end{eqnarray*}

The static, spherically symmetric metric which satisfies these
vacuum equations of fields for n-dimensional was obtained by
Boulware and Deser~\cite{BD} and in standard spherical coordinates
$(t,r,\theta_1 ,\cdots,\theta_{n-2})$ reads
\begin{equation}
ds^2=-f(r,t)dt^2+f^{-1}(r,t)dr^2
+r^2d\Omega^2_{n-2} \; ,
\end{equation}
where $f(r,t)$ is given by
\begin{equation}
f_{\pm}(r,v)= 1+\frac{r^2}{2\widehat{\alpha} }\left \{1\pm
\sqrt{1+\frac{8\widehat{\alpha}}{n-2}\left
[\frac{\Lambda}{n-1}+\frac{2M}{r^{n-1}}\right ]}\right \},\\
\end{equation}
with $\widehat{\alpha}=\alpha (n-3)(n-4)$ and $d\Omega^2_{n-2}$
being the metric of the $(n-2)$-sphere
\begin{equation}
d\Omega^2_{n-2} = d\theta^2_1 + \sum^{n-2}_{i=2}\prod^{i-1}_{j=1}
\sin^{2}\theta_j\;d\theta^2_i \; ,
\end{equation}
It is straightforward to prove that only the minus-branch
solution, $f_{-}(r,v)$, has limit for $\alpha\rightarrow 0$,
namely  n-dimensional GR.

In the next section we will find that this metric is a particular
case of a large family of  metrics. Although it is not crucial for
the proof of the theorem, we adopt $\alpha$ positive since this
condition arises from the strings theory. Hereafter we will assume
that both $\Lambda$ and $\alpha$ are fixed quantities.

\section{Radiating black holes in Einstein-Gauss-Bonnet gravity}
\textbf{Theorem 1:} \textit{Let $(\mathcal{M},g_{ab})$ a
n-dimensional space-time such that: i) it satisfies the EGB
equations, ii) it is spherically symmetric, iii) In the
Eddington-Bondi coordinates, where the metric reads
$\,ds^2=-h^2(r,v)f(r,v)dv^2+2\epsilon h(r,v)dvdr+r^2d\Omega^2_{n-2}$,
the energy-momentum tensor $T^a_b$ satisfies the conditions
$T^v_r=0$, and $T^{\theta_1}_{\theta_1}=kT^r_r$,
$(k=\text{const}\in \mathbb{R})$ iv) If $\alpha\rightarrow 0$, the
solution converges to the General Relativity limit. Then the
metric of the space-time is given by
\begin{eqnarray} ds^2=-f(r,v)dv^2+2\epsilon
dvdr+r^2d\Omega^2_{n-2},\,~~~~~~(\epsilon=\pm1),\label{metrica}\end{eqnarray}
where}
\begin{widetext}
\begin{equation}
f(r,v)=\left\{
\begin{array}{rll}
  1+\frac{r^2}{2\widehat{\alpha}}\left \{1-
\sqrt{1+\frac{8\widehat{\alpha}}{n-2}\left
[\frac{\Lambda}{n-1}+\frac{2M(v)}{r^{n-1}}-\frac{8\pi
C(v)}{[k(n-2)+1]r^{(1-k)(n-2)}}\right ]}\right \}\;&\text{if}&\;\; k\neq-\frac{1}{n-2},\\
  1+\frac{r^2}{2\widehat{\alpha}}\left \{1-
\sqrt{1+\frac{8\widehat{\alpha}}{n-2}\left
[\frac{\Lambda}{n-1}+\frac{2M(v)}{r^{n-1}}-\frac{8\pi
C(v)\ln(r)}{r^{n-1}}\right ]}\right \} \;\;\;\;~\;~~~\;\;\;&\text{if}&\;\; k=-\frac{1}{n-2}.\\
\end{array}\\\right.\label{lametrica}
\end{equation}
\end{widetext}
with the diagonal components of $T^a_b$ given by \begin{equation}
T^a_{b\text{(Diag)}}=\frac{C(v)}{r^{(n-2)(1-k)}}\text{diag}[1,1,k,\cdots,k],\label{Trr}
\end{equation}
and an unique non-vanishing off-diagonal element
\begin{equation} T^r_v=\left\{
\begin{array}{lll}
\frac{1}{4\pi
r^{n-2}}\frac{dM}{dv}-\frac{r^{(k-1)(n-2)+1}}{k(n-2)+1}\frac{dC}{dv}\;\;
&\text{if}&\;\; k\neq -\frac{1}{n-2}\\
\frac{1}{4\pi
r^{n-2}}\frac{dM}{dv}-\frac{\ln(r)}{r^{n-2}}\frac{dC}{dv} \;\;
&\text{if}&\;\; k= -\frac{1}{n-2}\\
\end{array}\\\right.
\end{equation}
with $M(v)\;,\;C(v)$ two arbitraries functions depending of the
distribution of the energy-matter.$\blacksquare$\\

\textbf{Proof:} By the hypothesis iii) of the theorem $1$, the
metric under discussion reads
\begin{equation}
ds^2=-h^2(r,v)f(r,v)dv^2+2\epsilon h(r,v)dvdr+r^2d\Omega^2_n,
\end{equation}
which due to hypothesis ii) must satisfy the EGB equations. The
only nontrivial components of the EGB tensor are
\begin{widetext}
\begin{eqnarray}
\mathcal{G}^v_r&=& (n-2)  \left [r^2+2( 1-f)\widehat{\alpha}
 \right ]\frac{h_r}{{r}^{3}
{\epsilon} h^{2}},\label{Gvr}\\
\nonumber\\
 \mathcal{G}^r_v&=&-(n-2)\left [ r^2+2(1-f)
\widehat{\alpha} \right ]\frac {f_v}{2{ r}^{3}},\label{Grv}\\
\nonumber\\
\mathcal{G}^r_r&=&\Lambda+(n-2)\left [ r^2+2( 1-f)
\widehat{\alpha}
 \right ]\frac{f_r}{2r^3}
 -(n-2)\left [ (n-3)r^2+( n-5)(1-f)\widehat{\alpha}
 \right ]\frac{1-f}{2r^4}\nonumber\\
 &&+(n-2)\left [ r^2+4(1-f) \widehat{\alpha}
 \right ]f\frac{h_r}{r^3h},\label{Grr}\\\nonumber\\
\mathcal{G}^v_v&=& \Lambda+(n-2)\left [ r^2+2( 1-f)
\widehat{\alpha}
 \right ]\frac{f_r}{2r^3}-(n-2)\left [(n-3) r^2+( n-5)(1-f)\widehat{\alpha}
 \right ]\frac{1-f}{2r^4},\label{Gvv}\\
 \nonumber\\
\mathcal{G}^{\theta_i}_{\theta_i}&=&
\Lambda-\frac{\widehat{\alpha}}{r^4}\left
[r^2f^2_r-2(n-6)(n-5)(1-f^2)\right ]
-(n-3)(n-4)\frac{1-f}{2r^2}\nonumber\\
 &&+\left [r^2+2\widehat{\alpha}(1-f)r^2\right ]\left [(fh^2h_{rr}+
 hh_{rv}-h^3 f_{rr})-\epsilon h_rh_v\right ]\frac{1}{r^2h^3}\nonumber\\
 &&+\left [(n-3)r^2+2(n-5)\widehat{\alpha}(1-f)\right
 ]\frac{f_r}{r^3}\nonumber\\
 &&+\left \{6r^3h+4\widehat{\alpha}\left [
(3-5f)rh+4(n-5)(1-f)hf-4\epsilon rf_v\right
]+(n-3)\frac{f}{r}\right
\}\frac{h_r}{4r^3h^2}.\label{Gtt}\end{eqnarray}
\end{widetext}

By using hypothesis $iii)$, $T^v_r=0$, in Eq.(\ref{Gvr}), we
deduce that: either $h(r,v)$ is independent of $r$, i.e.,
$$h(v,r)=h(v),$$ or $f(v,r)$ is independent of $v$, and
satisfies \begin{equation} r^2+2(
1-f)\widehat{\alpha}=0.\label{fotrocaso}\end{equation}

However, this last equation leads to non-radiating metrics (
because Eq.(\ref{Grv}) implies $T^r_v=0$), which are isometric to
AdS/dS Gauss-Bonnet solutions, and naturally contained in the
metrics Eq.(\ref{lametrica}) as particular cases with
$M(v)=C(v)=0$. We show this in the Appendix A. Then, we conclude
that $h(r,v)=h(v)$, and redefining to the null coordinate
$\widehat{v}=\int h(v)dv,$ we can take $h(v)=1$, without loss of
generality.

Now, from Eq.(\ref{Grr}) and Eq.(\ref{Gvv}), we obtain that
$G^v_v=G^r_r$, and then $$T^v_v=T^r_r.$$

If we impose the conservation laws, $\nabla_a T^a_b=0,$ and using
again the hypothesis $iii)$, $T^{\theta_i}_{\theta_i}=kT^r_r$, we
have that
\begin{eqnarray}
0&=&\frac{\partial }{\partial v}{ T^v_v} +\frac {\partial
}{\partial r} T^r_v +\frac{1}{2\epsilon}\left (T^r_r- T^v_v\right
){\frac
{\partial }{\partial r}f}+ \frac {( n-2)}{r}T^r_v,\nonumber\\
&&\label{Bt11}\\
0&=&\frac {\partial }{\partial r}T^r_r +\frac {( n-2)(1-k)}
{r}T^r_r. \label{Bt22}
\end{eqnarray}

Solving Eq.(\ref{Bt22}) for $T^r_r$, we obtain:
\begin{equation}
T^r_r=\frac{C(v)}{r^{(n-2)(1-k)}},\label{Trr}
\end{equation}
where $C(v)$ is an arbitrary function.

By joining these results, we can write the diagonal elements of
$T^a_b$ as
$$ T^a_{b\text{(Diag)}}=\frac{C(v)}{r^{(n-2)(1-k)}}\text{diag}[1,1,k,\cdots,k].$$

Now, we can find $f(r,v)$ by replacing Eq.(\ref{Grr}) and
Eq.(\ref{Trr}) in the suitable component of the EGB equations
$$\mathcal{G}^r_r=8\pi T^r_r,$$ resulting after solving this differential equation and making some
algebraic simplifications
\begin{widetext}
\begin{equation}
f_{\pm}(r,v)=\left\{
\begin{array}{rll}
  1+\frac{r^2}{2\widehat{\alpha}}\left \{1\pm
\sqrt{1+\frac{8\widehat{\alpha}}{n-2}\left
[\frac{\Lambda}{n-1}+\frac{2M(v)}{r^{n-1}}-\frac{8\pi
C(v)}{[k(n-2)+1]r^{(1-k)(n-2)}}\right ]}\right \}\;&\text{if}&\;\; k\neq-\frac{1}{n-2},\\
  1+\frac{r^2}{2\widehat{\alpha}}\left \{1\pm
\sqrt{1+\frac{8\widehat{\alpha}}{n-2}\left
[\frac{\Lambda}{n-1}+\frac{2M(v)}{r^{n-1}}-\frac{8\pi
C(v)\ln(r)}{r^{n-1}}\right ]}\right \} \;\;\;\;~\;~~~\;\;\;&\text{if}&\;\; k=-\frac{1}{n-2}.\\
\end{array}\\\right.
\end{equation}
\end{widetext}
where $M(v)$ is another arbitrary function (which can be shown to
be proportional to the mass of the underlying matter ).

At this point, it is important to note that we have obtained two
branches of solutions, namely, $f_+$ and $f_-$, which correspond to $\pm$ signs
in front of the square root term. However, the positive branch, $f_+$,
does not converge to GR.

Then, if we impose the hypothesis $iv)$, then it can be shown that
the only possible solutions are the following:
\begin{widetext}
\begin{equation}
f(r,v)=\left\{
\begin{array}{lll}
  1+\frac{r^2}{2\widehat{\alpha}}\left \{1-
\sqrt{1+\frac{8\widehat{\alpha}}{n-2}\left
[\frac{\Lambda}{n-1}+\frac{2M(v)}{r^{n-1}}-\frac{8\pi
C(v)}{[k(n-2)+1]r^{(1-k)(n-2)}}\right ]}\right \}\;&\text{if}&\;\; k\neq-\frac{1}{n-2},\\
  1+\frac{r^2}{2\widehat{\alpha}}\left \{1-
\sqrt{1+\frac{8\widehat{\alpha}}{n-2}\left
[\frac{\Lambda}{n-1}+\frac{2M(v)}{r^{n-1}}-\frac{8\pi
C(v)\ln(r)}{r^{n-1}}\right ]}\right \} \;\;\;\;~\;~~~\;\;\;&\text{if}&\;\; k=-\frac{1}{n-2}.\\
\end{array}\\\right.
\end{equation}
\end{widetext}
In this case, the limit $\alpha\rightarrow 0$ reduces $f(r,v)$ to
\begin{widetext}
\begin{equation}
f(r,v)=\left\{
\begin{array}{lll}
1-\frac{4M(v)}{(n-2)r^{n-3}}-\frac{2\Lambda
r^2}{(n-1)(n-2)}+\frac{16\pi
C(v)}{(n-2)[k(n-2)+1]r^{(1-k)(n-2)-2}}\;&\text{if}&\;\; k\neq-\frac{1}{n-2},\\\\
1-\frac{4M(v)}{(n-2)r^{n-3}}-\frac{2\Lambda
r^2}{(n-1)(n-2)}+\frac{16\pi C(v) ln(r)}{(n-2)r^{n-3}}\;\;\;\;~\;~~~\;\;\;&\text{if}&\;\; k=-\frac{1}{n-2}.\\\\
\end{array}\\\right.
\end{equation}
\end{widetext}

These are the expressions that we would have found if we had
started with the $n$-dimensional Einstein version of the theorem
(General Relativity with a cosmological constant) instead of the
Einstein-Gauss-Bonnet gravity. In fact, if $n=4$, these
expressions are the same that those found in~\cite{Ghosh}.

Finally, from Eq.(\ref{Grv}) and the appropriated EGB equation,
$$\mathcal{G}^r_v=8\pi T^r_v,$$
we obtain that the only non-vanishing off-diagonal el\-ement of
$T^a_b,$ reads
\begin{equation}
T^r_v=\left\{
\begin{array}{lll}
\frac{1}{4\pi
r^{n-2}}\frac{dM}{dv}-\frac{r^{(k-1)(n-2)+1}}{k(n-2)+1}\frac{dC}{dv}\;\;
&\text{if}&\;\; k\neq -\frac{1}{n-2}\\
\frac{1}{4\pi
r^{n-2}}\frac{dM}{dv}-\frac{\ln(r)}{r^{n-2}}\frac{dC}{dv} \;\;
&\text{if}&\;\; k= -\frac{1}{n-2}\\
\end{array}\\\right.
\end{equation}
Q.E.D.\\\\

This family of metrics contains many potentially interesting
solutions. Some of these space-times are shown in the Table 1.
These solutions could be very useful to study the collapse of
different matter fields, or the formation of naked singularities.
On the other hand, we also think that they can be used to analyzed
semiclassical approaches to the evaporation of black holes.

However, the imposed conditions on the energy-momentum tensor are
rather restrictive so other important solutions are not allowed.
In the section V we generalize the theorem $1$ in order to get new
exact solutions.

\begin{table*}
\caption{\label{tab:table3}Some space-times satisfying the
conditions of the Theorem $1$ and $2$, obtained with particular
values of $k$, $C(v)$ and $M(v).$ }
\begin{ruledtabular}
\begin{tabular}{llll}

 $T^a_b$&Space-Time&M(v) and C(v) &k-index\\ \hline\\
 $T^a_{b\text{(Diag)}}=0\;,\;T^r_v=\frac{1}{4\pi r^{n-2}}\frac{dM}{dv}$&Kobayashi-Maeda&$C(v)=0$
 &\\\\\hline\\
 $T^a_{b\text{(Diag)}}=-\frac{Q^2(v)}{4\pi r^{n-2}}\text{diag}[1,1,-1,\cdots,-1]$&Bonnor-Vaidya-EGB
 &$C(v)=-\frac{Q^2(v)}{8\pi}$&$k=-1$\\\\ $T^r_v=\frac{1}{4\pi
r^{2n-5}}\left[r^{n-3}\frac{dM}{dv}-\frac{Q}{(n-3)}\frac{dQ}{dv}\right]$&&&
\\\\\hline\\
 $T^a_{b\text{(Diag)}}=-\frac{g^2(v)}{4\pi r^{(n-2)(m+1)}}\text{diag}[1,1,-m,\cdots,-m]$&Husain-EGB&$C(v)=
 -\frac{g^2(v)}{4\pi}$&$k=-m$\\\\ $ T^r_v=\frac{1}{4\pi}
 \left[r^{2-n}\frac{dM}{dv}-r^{(m+1)(2-n)+1}\frac{2g}{1-m(n-2)}\frac{dg}{dv}\right]$&&&
 \\\\\hline\\
 $ T^v_v=T^r_r=-\frac{a}{8\pi r^{n-2}}$& Global monopole-EGB
 &$M(v)=0, \;\;C(v)=-\frac{a}{4\pi}$&$k=0$\\\\\hline\\
 $T^a_b=0$&Boulware-Deser-Wheeler
&$M(v)=M_0$, $C(v)=0$&
 \\\\\hline\\
$T^r_r=T^v_v=\frac{Q^2}{4\pi
L^2r^{n-2}}\left(r^{n-2}-\sqrt{r^{2n-4}+L^2}\right)$&BI-Vaidya-EGB&$C(v)=-\frac{Q^2(v)}{4\pi
|L|}$& $k(r,v)=-\frac{r^{n-2}}{\sqrt{r^{2(n-2)}+L^2}}$\\\\
$T^{\theta_i}_{\theta_i}=k(r,v)T^r_r$&&&
 \\\\
 $T^r_v=\frac{1}{4\pi r^{n-2}}\left[\frac{dM}{dv}-br\left(\frac{dQ}{dv}F+Q\frac{\partial F}{\partial v}\right)\right]$&&&\\
\end{tabular}
\end{ruledtabular}
\end{table*}

\section{Energy conditions}
In this section we discuss the restrictions on the energy-momentum
tensor based in the dominant energy conditions. These will impose
restrictions on $k$, $M(v)$ and $C(v)$, defined in the last
section. The case of weak energy conditions is totally analogous
to that discussed in~\cite{Ghosh} for General Relativity theory.

The covariant components of the energy-momentum tensor can be
written with the help of two future null vectors, $l_a$, $n_a$
(the vector $l^a$ is tangent to the null surface generated by $v$,
and $n^a$ is an independent null vector such as $l_an^a=-1$; in
Eddington-Bondi coordinates $\{v,r,\theta_i\}$, the components of
these vectors are $l_a=(1,0,\cdots,0)$, $n_a=(f/2,-1,0,\cdots,0)$)
as follows,
$$T_{ab}=\epsilon \mu
l_al_b-P_r(l_an_b+l_bn_a)+P_{\theta}(g_{ab}+l_an_b+l_bn_a),$$
where
\begin{eqnarray}
\mu&=&T^r_v\\
P_r&=&-T^r_r=-C(v)r^{(2-n)(1-k)}\\
P_{\theta}&=&kP_r.
\end{eqnarray}

Physically $\mu$ represents the radiating energy along the null
direction given by $l^a$; $P_r$ denotes the radial pressure
generated for the charges of the fluids and $P_{\theta}$ are the
transversal pressures. All these quantities are measured by an
observer moving along a time-like direction $u^a$ given by
$$u^a=\frac{1}{\sqrt{2}}(l^a+n^a).$$ This observer will measure an
energy density given by
$$\rho=-P_r=-T_{ab}u^au^b.$$ Note that the energy-momentum tensor
corresponds to a null fluid Type II.

As it is well known, the dominant energy conditions implies that
for all timelike vector $t^a$, $T_{ab}t^at^b\geq 0$ and also
$T_{ab}t^b$ is non-spacelike vector. Then, we must require:

$$\mu\geq 0\;~~~~~~\; \text{and} \;~~~~~~~\rho\geq P_{\theta}\geq 0.$$

The first condition, in the case of a radiating fluid $\mu>0$, is
equivalent to
\begin{equation}
\frac{1}{4\pi
r^{n-2}}\frac{dM}{dv}-\frac{r^{(k-1)(n-2)+1}}{k(n-2)+1}\frac{dC}{dv}>
0\;\; \text{if}\;\; k\neq
-\frac{1}{n-2},\label{conkdis}\end{equation} and
\begin{equation}
\frac{1}{4\pi
r^{n-2}}\frac{dM}{dv}-\frac{\ln(r)}{r^{n-2}}\frac{dC}{dv}> 0
\;\;\text{if}\;\; k= -\frac{1}{n-2}.\label{conkigual}
\end{equation}
The Eq.(\ref{conkdis}) is satisfied if $\frac{dM}{dv}>0$, and
either $\frac{dC}{dv}>0$ with $k< -\frac{1}{n-2},$ or
$\frac{dC}{dv}<0$ with $k> -\frac{1}{n-2}.$

On the other hand, the Eq.(\ref{conkigual}) is satisfied if
$\frac{dM}{dv}>4\pi \ln(r)\frac{dC}{dv}.$

Finally, the conditions $\rho\geq P_{\theta}\geq 0$ are satisfied
only if $C(v)\leq 0$ and $-1\leq k \leq 0$.

\section{A more general version of the Theorem}
In this section, we present a slightly different version of the
theorem, which allows more general distributions of matter fields.
\\\\
\textbf{Theorem 2:}\textit{ Let $(\mathcal{M},g_{ab})$ a
n-dimensional space-time such that: i) it satisfies the EGB
equations, ii) it is spherically symmetric, iii) In the
Eddington-Bondi coordinates, where the metric reads
$\,ds^2=-h^2(r,v)f(r,v)dv^2+2\epsilon h(r,v)dvdr+r^2d\Omega^2_n$,
the energy-momentum tensor $T^a_b$ satisfies the conditions
$T^v_r=0$, and $T^{\theta_1}_{\theta_1}=k(r,v)T^r_r$, with
$k(r,v)$ a real function, iv) If $\alpha\rightarrow 0$, then the
solution converges to the General Relativity limit. Then the
metric of the space-time is given by
$$ds^2=-f(r,v)dv^2+2\epsilon
dvdr+r^2d\Omega^2_n,\,~~~~~~(\epsilon=\pm1),$$ where
\begin{widetext}
\begin{equation}
f(r,v)=
  1+\frac{r^2}{2\widehat{\alpha}}\left \{1-
\sqrt{1+\frac{8\widehat{\alpha}}{n-2}\left
[\frac{\Lambda}{n-1}+\frac{2M(v)}{r^{n-1}}-\frac{8\pi
C(v)}{r^{n-1}}I(r,v)\right ]}\right \}
\end{equation}
\end{widetext}
with
\begin{equation}
I(r,v)=\int^r_0{e^{(n-2)\int^s{k(t,v)t^{-1}}dt}ds},
\end{equation}
and the diagonal components of $T^a_b$ given by
\begin{equation}
T^a_{b\text{(Diag)}}=\frac{C(v)e^{(n-2)\int{k(r,v)r^{-1}}dr}}{r^{(n-2)}}\text{\textrm{diag}}[1,1,k,\cdots,k],
\end{equation}
with an unique non-vanishing off-diagonal element
\begin{equation}
T^r_v=\frac{1}{
r^{n-2}}\left[\frac{1}{4\pi}\frac{dM}{dv}-\frac{dC}{dv}I(r,v)-C\frac{\partial
I}{\partial v}\right ]
\end{equation}
where $M(v)\;,\;C(v)$ are two arbitraries functions depending of
the distribution of the energy-matter.}$\blacksquare$\\

The proof is completely analogous to that of the previous theorem.
Note that if $k=\text{const},$ we recover the results of that
Theorem. In the next example, we apply this theorem in order to
obtain the metric of a null charged fluid in Born-Infeld nonlinear
electrodynamics.
 \section{An example: The radiating Born-Infeld black
hole} Since Thompson discovered the electron, the physicists were
very worried about the divergence of the self-energy of point-like
charges in the Maxwell electrodynamics. There were a lot of
proposals to solve this problem, and one of the most prominent
theories compatible with the spirit of Lorentz invariance was
created by Born and Infeld between 1932 and 1934~\cite{BI1}. They
proposed a nonlinear and Lorentz invariant Lagrangian, which depends
of one parameter $b$, such as, if $b\rightarrow\infty$, then its
Lagrangian tends to the Maxwell Lagrangian. The electromagnetic
field $F_{ab}$ can be obtained from a 1-form $A_a$ in a similar way
to Maxwell theory, $F=dA$. In the last years, the interest in this
theory was renewed due to the fact that it appears as a low energy
limit of string theory~\cite{cuerdas1,cuerdas2}.

The free Lorentz invariant Lagrangian suggested by Born-Infeld
reads
$$\frak{L}=\frac{b^2\sqrt{-g}}{4\pi}\left (1-\sqrt{1+\frac{F^{ab}F_{ab}}{2b^2}}\right
),$$ where $g$ is the determinant of the metric $g_{ab}$. If we
add an interaction term in the form of $J^aA_a$ to the Born-Infeld
Lagrangian, then the equations of motions that follows from
$\frak{L}$ are
\begin{equation}
\frac{1}{\sqrt{-g}}\frac{\partial}{\partial{x^a}}\left
(\frac{\sqrt{-g}F^{ab}}{\sqrt{1+\frac{F^{ab}F_{ab}}{2b^2}}}\right
)=4\pi J^b.
\end{equation}
In addition to the \textit{Bianchi identities}
$$F_{ab,c}+F_{ca,b}+F_{bc,a}=0.$$
Now, we are interested in finding a spherically symmetric
Vaidya-like solution with a charged null fluid in Born-Infeld
electrodynamics.

Let us consider a spherically symmetric and non-stationary
electromagnetic field of the form
$$F=E(r,v)dv\wedge dr.$$ In this case, the contravariant
components of $F_{ab}$ in the metric $g_{ab}$ (
Eq.(\ref{metrica})) are
\begin{eqnarray}
F^{rv}&=&-\epsilon F_{rv}=\epsilon E(r,v)\\
F^{vr}&=&\epsilon F_{rv}=-\epsilon E(r,v)
\end{eqnarray}
and the determinant of the metric reads
\begin{equation}
g=-r^{2(n-2)}\prod^{n-2}_{i=3}\left[\sin(\theta_{i})\right]^{n-i}.
\end{equation}
Note, because we are considering a null fluid, there will be a
null current given by
\begin{equation}
J^a=J(r,v)\delta^a_v.
\end{equation}
Having all these expressions into account, the equations of
motions for $F_{ab}$ are,
\begin{eqnarray}
\frac{1}{\sqrt{-g}}\frac{\partial}{\partial{r}}\left
(\frac{\sqrt{-g}E(r,v)}{\sqrt{1-\frac{E^2(r,v)}{b^2}}}\right
)&=&0,\\
\frac{1}{\sqrt{-g}}\frac{\partial}{\partial{v}}\left
(\frac{\sqrt{-g}E(r,v)}{\sqrt{1-\frac{E^2(r,v)}{b^2}}}\right
)&=&4\pi  J(r,v).
\end{eqnarray}
Solving this equations we find
\begin{equation}
E(r,v)=\frac{Q(v)}{\sqrt{r^{2(n-2)}+L^2(v)}}
\end{equation}
where
\begin{equation}
L(v)=\frac{Q(v)}{b},
\end{equation}
and $Q(v)$ an arbitrary function of $v$, representing the charge
of the fluid. The null current results, $$J^a=\frac{1}{4\pi
r^{n-2}}\frac{dQ}{dv}\delta^a_v.$$ Note that if
$b\rightarrow\infty$ then $E(v,r)$ tends to its Maxwell
expression. Note also that the 1-form $A_a$ from this problem has
the form:
$$A=\left[\int_0^r{\frac{dr}{\sqrt{{r^{2(n-2)}}+L^2}}}\right]Q(v)dv.$$

The energy-momentum tensor associated to the field $F_{ab}$ is,
$$T_{ab}=\frac{1}{4\pi}\left(\frac{b^2g_{ab}-F_{ad}F_b\,^d-\frac{1}{2}g_{ab}F_{cd}F^{cd}}{\sqrt{1+\frac{1}{2b^2}F_{cd}F^{cd}}}-b^2g_{ab}\right).$$
From this and from the form of the current, the non-radiative part
of the energy-momentum tensor of the charged null fluid results
\begin{eqnarray}
T^r_r&=&\frac{Q^2}{4\pi
L^2r^{n-2}}\left(r^{n-2}-\sqrt{r^{2(n-2)}+L^2}\right),\\\nonumber\\
T^v_v&=&T^r_r,\\
T^{\theta_i}_{\theta_i}&=&-\frac{r^{n-2}}{\sqrt{r^{2(n-2)}+L^2}}T^r_r,
\end{eqnarray}

and the other off-diagonal components vanishes (with the exception
to $T^r_v$).

Then, we can see that this null charged fluid satisfies the
conditions of the Theorem 2, with
$$k(r,v)=-\frac{r^{n-2}}{\sqrt{r^{2(n-2)}+L^2}},$$ and
$$C(v)=-\frac{Q^2}{4\pi |L|}=-\frac{|bQ(v)|}{4\pi},$$ because it can be easily computed that
$$e^{(n-2)\int{k(r,v)r^{-1}dr}}=-\frac{1}{|L|}\left(r^{n-2}-\sqrt{r^{2(n-2)}+L^2}\right).$$

If we replace these expressions of $k(r,v)$ and $C(v)$ into
$f(r,v)$, we obtain the metric for a Born-Infeld dynamic black
hole:
\begin{widetext}
\begin{equation}
f(r,v)=
  1+\frac{r^2}{2\widehat{\alpha}}\left \{1-
\sqrt{1+\frac{8\widehat{\alpha}}{n-2}\left
\{\frac{\Lambda}{n-1}+\frac{2M(v)}{r^{n-1}}-\frac{2
b^2}{(n-1)}+\frac{2
b^2}{r^{n-1}}\int^r_0{\sqrt{r^{2(n-2)}+L^2}dr}\right ]}\right \}
\end{equation}
\end{widetext}
In the case of $Q(v)=Q_0$, $M(v)=M_0$, we recover the static
Born-Infeld solution, first obtained by Wiltshire~\cite{WSH1} and
reexamined in~\cite{Giribet}.

It can be shown that the integral factor
$$\mathcal{I}=\int^r_0{\sqrt{r^{2(n-2)}+L^2}dr},$$ can be written
in terms of a generalized hypergeometric function
$F([a_i],[b_j],z)$ (dependent of $p$ parameters $a_i$ and $q$
parameters $b_j$) defined by
$$F([a_i],[b_j],z)=\sum _{k=0}^{\infty}\prod _{i=1}^{p}\prod _{j=1
}^{q}{\frac {\Gamma  \left( a_{{i}}+k
 \right)\Gamma(b_j) }{\Gamma  \left( a_{{i}} \right){\Gamma  \left( b_{{j}}+k
\right) }}}\frac{{z}^{k}}{k!}.$$

 The result is:
 \begin{equation}\ \mathcal{I}= |L|\,r F\left( \left[-\frac{1}{2}, \frac{1}{2n-4}\right],\left[
 \frac{2n-3}{2n-4}\right],-{\frac {r^{2(n-2)}}{L^2}}
 \right).\label{hyp}\end{equation}

 Finally, if we compute the $T^r_v$ component of the
 energy-momentum tensor using the theorem $2$, we get
$$T^r_v=\frac{1}{4\pi r^{n-2}}\left[\frac{dM}{dv}-br\left(\frac{dQ}{dv}F+Q\frac{\partial F}{\partial v}\right)\right],$$
where $F$ denotes the hypergeometric function defined in
Eq.(\ref{hyp}).
\section{Conclusions}
In this work, we have proved two theorems, which allow us to
obtain exact solutions to the Einstein-Gauss-Bonnet equations.
This solutions represent dynamic black holes, and we thought that
these solutions are of interest, because they generalize some
known solutions of GR to the context to Einstein-Gauss-Bonnet
gravity which has been proved, is the first nontrivial term of low
energy limit of string theory.

It is important to note that with plus sign in front of the square
root, there are black hole solution even when the
$(n-2)$-dimensional submanifold is plane or hyperbolically
symmetric. The analogous theorem seems to be hold in these cases.

In a future work, we will make a detailed study of these metrics,
analyzing the causal structure and its thermodynamics.

Also, it should be very interesting to apply these metrics to
study the matter collapse, the naked singularities formations, and
semiclassical analysis of evaporation of black holes. These works,
are now in progress.

\begin{acknowledgments}
E.G.  would like to thank CONICET for support. A. E. D. would like
to thank Instituto Universitario Aeron\'autico for support. We
also thank an anonymous referee for their interesting comments and
suggestions which helped us to improve this work.
\end{acknowledgments}

\appendix

\section{}
In this appendix, we will analyze the consequences to impose the
Eq.(\ref{fotrocaso}) as solution to the Eq.(\ref{Gvr}).

We will use the conservation laws, $\nabla_a T^a_b=0$, which, as
we saw, implies that
\begin{eqnarray}
T^r_r&=&C(v)r^{(-1+k)(n-2)},\label{trroc}\\
T^{\theta_i}_{\theta_i}&=&k T^r_r=kC(v)r^{(1-k)(n-2)}.\label{ttt}
\end{eqnarray}

The solution of the Eq.(\ref{fotrocaso}) reads,
\begin{equation}
f(r,v)=1+\frac{r^2}{2\widehat{\alpha}}.\label{fotrosol}
\end{equation}
 By replacing Eqs.(\ref{trroc}), (\ref{ttt}) and (\ref{fotrosol}) in the EGB equations:

\begin{eqnarray}
&\mathcal{G}^r_v-8\pi T^r_v&=0,\\
&\mathcal{G}^v_v-8\pi T^v_v&=0,\\
&\mathcal{G}^{\theta_i}_{\theta_i}-8\pi
T^{\theta_i}_{\theta_i}&=0,
\end{eqnarray}
we get respectively
\begin{eqnarray}
&&T^r_v=0,\\
&&T^v_v=\frac{\Lambda}{8\pi}+\frac{(n-1)(n-2)}{64\pi
\widehat{\alpha}},
\end{eqnarray}
and
\begin{equation}\Lambda+\frac{(n-1)(n-2)}{8\widehat{\alpha}}-8\pi
kC(v)r^{(1-k)(n-2)}=0.\label{posiblec}
\end{equation}

Now, Eq.(\ref{posiblec}) has two possible solutions:

\underline{Case 1:} $C(v)=0.$ In this case, the
Eq.(\ref{posiblec}) can be satisfied if and only if the constants
$\Lambda$ and $\widehat{\alpha}$ are related in the following
form:
\begin{equation}
\Lambda=-\frac{(n-1)(n-2)}{8\widehat{\alpha}},
\end{equation}
which, when is replaced in the r-r component of the EGB equations,
$$\mathcal{G}^r_r-8\pi T^r_r=0,$$ gives a differential equation
for $h(r,v)$:
\begin{equation}(n-2)(1+\frac{r^2}{2\widehat{\alpha}})\frac{h_r}{rh}=0.\label{ecparah}\end{equation}
This implies again, that $h(r,v)=h(v)$, which can be taken as 1 by
an appropriated coordinate transformation.

Finally, collecting the previous results, we obtain that the
energy-momentum tensor is zero, $$T^a_b=0.$$

The resulting metric, is given by:
$$\,ds^2=-\left(1+\frac{r^2}{2\widehat{\alpha}}\right )dv^2+2\epsilon dvdr+r^2d\Omega^2_n$$
which are the known solutions de Sitter/anti-de Sitter (dS/AdS) metrics in EBG gravity.\\\\
\underline{Case 2:} $C(v)=\text{const},$ and $k=1$. In this case,
from Eq.(\ref{posiblec}) we get that
\begin{equation}
C(v)=\frac{\Lambda}{8\pi}+\frac{(n-1)(n-2)}{64\pi
\widehat{\alpha}},
\end{equation}
which, when it is replaced in the r-r component of the EGB
equations, $$\mathcal{G}^r_r-8\pi T^r_r=0,$$ gives the same
differential equation for $h(r,v)$ that in the Case 1,
Eq.(\ref{ecparah}), and again we can take $h(r,v)=1$.

By replacing the expressions of $C(v)$ and $f(r)$ in the EGB
equations, we conclude that the only non-zero components of
$T^a_b$ are in its diagonal part,
\begin{equation}
T^a_b=\left [\frac{\Lambda}{8\pi}+\frac{(n-1)(n-2)}{64\pi
\widehat{\alpha}}\right ]\delta^a_b.
\end{equation}

This energy-momentum tensor, can be thought as an effective vacuum
energy, and from this, resulting a cosmological effective term
$\widehat{\Lambda}$ given by
$$\widehat{\Lambda}=-\frac{(n-1)(n-2)}{8\widehat{\alpha}}.$$

These metrics are again dS/AdS solutions to the EGB equations.
Note that these two cases are contained in the expression of the
metric in the theorem, if we take $M(v)=C(v)=0$.

\end{document}